# Whipped and mixed warm clouds in the deep sea

**by Hans van Haren*[,ar], KM3NeT collaboration*[,]****

*Corresponding author: hans.van.haren@nioz.nl; km3net-pc@km3net.de

[ar]NIOZ (Royal Netherlands Institute for Sea Research), PO Box 59, Den Burg, Texel, 1790 AB, the Netherlands

**Complete author list can be found in the Appendix

**Key points:**

- Three-dimensional large mooring array in the deep Northwestern Mediterranean held nearly 3000 high-resolution temperature sensors
- The moored observations are cast in short 3D movies revealing development of turbulent convection in the deep sea
- Three types of turbulence generation are quantified: slantwise-convected warm-water clouds, geothermal heating, interfacial-wave breakdown

**Abstract** Turbulence is indispensable to redistribute nutrients for all life forms larger than microbial, on land and in the ocean. Yet, the development of deep-sea turbulence was not studied in three dimensions to date. As a disproportionate laboratory, an array of nearly 3000 high-resolution temperature sensors had been installed for three years on the flat 2500-m deep bottom of the Mediterranean Sea. The time series from the half-cubic hectometer mooring-array allows for the creation of unique movies of deep-sea water motions. Although temperature differences are typically 0.001°C, variable convection-turbulence is observed as expected from geothermal heating through the flat seafloor. During about 40% of the time, an additional turbulence, 3 times stronger in magnitude, is observed from slantwise advected warmer waters to pass in turbulent clouds. Besides turbulent clouds and seafloor heating, movies also reveal weakly turbulent interfacial-wave breakdown that commonly occurs in the open ocean far away from boundaries.

**Plain Language Summary** Without deep-sea turbulence, life as we know it would not exist. Developments of deep-sea turbulent motions were studied using a three-dimensional array of nearly 3000 high-resolution temperature sensors. The array was installed on the flat 2500-m deep bottom of the Mediterranean Sea for three years. The temperature data allow for the creation of unique movies of deep-sea water motions. The movies show 0.0001°C temperature differences in "flames" of up to 100-m tall due to geothermal heating through the seafloor. During about 40% of the time, 3 times stronger turbulence is carried by warmer-water clouds from above and sideways. Besides turbulent clouds and seafloor heating, movies also reveal 3-10 times weaker small internal-wave breaking that commonly occurs in the open ocean far away from boundaries.

## 1. Introduction

Below the wind-agitated sea surface, the sea waters are not quiescent but also constantly in motion. These ocean-interior motions occur on a wide variety of scales, from several hundred kilometres large-scale circulations via tens of kilometres 'mesoscale' eddies to tens of meters turbulent overturns that eventually lose their energy at millimeter scales. One of the major contributors to ocean-interior motions are internal waves, which are largely driven by stable vertical density stratification of generally warm, fresh waters above cooler, saltier waters. In contrast to waves at the ocean surface, internal waves propagate in three dimensions (3D) including into the deep ocean (LeBlond & Mysak, 1978). Major sources of internal waves are tides and (near) inertial motions. The latter are remainders from the passage of atmospheric disturbances, deflected by the Earth's rotation. Commonly, freely propagating internal waves can exist between the inertial frequency and the buoyancy frequency, which is related to the stratification. Internal waves transfer their energy to irreversible turbulence when they deform and break. Deep-sea turbulence is mostly generated by internal-wave breaking at steep seafloor topography (Eriksen, 1982; Thorpe, 1987).

Turbulence is of high interest as it is an enabling factor for life for the spread of oxygen and nutrients. However, precise development of turbulence in 3D is difficult to study, one reason being that it extends over many orders of magnitude. In the deep sea, turbulence is commonly studied using instrumentation attached to a single line that is lowered from a ship or moored at the seafloor, thereby limiting the observations to 1D.

In this paper, year-long observations are presented from a custom-made, 45-line mooring-array with densely spaced high-resolution temperature 'T'-sensors to monitor nearly half-a-million cubic meters of deep-sea waters. The array was deployed in the Western Mediterranean at a flat seafloor of 2458-m water depth, in the vicinity of the under-water cubic kilometre neutrino telescope KM3NeT/ORCA off France (Adrián-Martinez, 2016). Tidal energy is low in this part of the Mediterranean, but alpine winds regularly induce near-inertial waves. These internal waves can combine with mesoscale eddies along with a distinct all-year round boundary waterflow, stronger in winter (Gascard, 1978; Crepon et al., 1982; Taupier-Letage & Millot, 1986; Testor & Gascard, 2006). Of the expected vertical transport processes in the area, deep dense-water formation due to the potentially highly-turbulent sinking of unstable cool surface waters is known to occur, but is rare at decadal scales (Mertens & Schott, 1998). Thus far, it has never been observed directly (Thorpe, 2005). However, eddies (Testor & Gascard, 2006) and near-inertial and high-frequency nonlinear internal waves (van Haren et al., 2014) that generate turbulence upon breaking may transport heat downward. This heat transport from above may mask turbulent geothermal heating (GH) through the seafloor from below. Previous shipborne observations have suggested that GH generates about twice as energetic turbulence than internal-wave breaking for the entire Northwestern Mediterranean (Ferron et al., 2017).

Materials and methods are in Supporting Information. Data from the 3D mooring-array allow for two major advancements in the study of turbulence: 1) Statistics of nearly 3000 independent T-sensors improve calculations of turbulence quantities and spectral information. This enhances understanding of coupling between turbulent motions and sources like internal waves and eddies. 2) Movies, created from the four-dimensional data enable studies of the development of various turbulence sources.

## 2. Results

In Fig. 1 the lower 500 m above the seafloor are shown of a standard oceanographic single CTD vertical (z) profile. It demonstrates a correspondence between Conservative Temperature (IOC et al., 2010), henceforth 'temperature' for short, and density variations. The relationship between variations in temperature ($\Theta$, Fig. 1a) and density anomaly referenced to 1000 kg m$^{-3}$ at a pressure of $2\times10^7$ N m$^{-2}$ ($\sigma_2$, Fig. 1b) is found $\delta\sigma_2/\delta\Theta = -0.25\pm0.01$ kg m$^{-3}$ °C$^{-1}$ over the stratified waters at $z>-2200$ m.

In these waters around $z=-2050$ m, the large-100-m scale buoyancy frequency amounts to $N\approx2f$, where f denotes the local inertial frequency (Fig. 1b). In the lower 250 m above seafloor, $N\approx(0.5\text{-}1)f$. As a result, the frequency range of freely propagating internal waves is about half the size of those in more stratified waters above. Over 1-10 m small vertical scales, larger $N\geq4f$ can be found, e.g. $z=-2110$ and $-1980$ m in Fig. 1. Such 1-10 m thin, stratified layers are occasionally also found at greater depths, as is best seen in Fig. 1a rather than in the noisier profile of Fig. 1b. Thin layers can carry interfacial waves. Thus, a considerable variation is seen in vertical stratification, but the situation is not stable in time as may be suggested from the CTD profile.

First-year time series of the data are shown in Fig. 2. Short periods of increasing temperature are frequently observed. As these data are from a single line, the direction of motion cannot be established. For this, a comparison with data from multiple lines has to be made. From a single line temperature differences can be observed of up to 0.015°C above background values for the uppermost T-sensors at h=125 m above seafloor and smaller values below (Fig. 2a). Such relatively warm stratified-water (SW) conditions alternate in time with near-homogeneous (NH) conditions of hardly any temperature difference over the 124-m T-sensor range. Also at the lowest T-sensor, periods of NH are generally found to be cooler than SW, with an occurrence in time of about 40% and 60% for SW and NH, respectively. Cool, substantially negatively stratified, unstable periods are not observed. The stratification variation contrasts with the near-bottom CTD-profile. Stratification increase either occurs via lowering warmer waters about 400 m vertically or advecting them horizontally. Internal waves and eddies are thought primary drivers of such motions that, combined, may lead to convection-turbulence that is slanted to the vertical (Marshall & Schott, 1999; Straneo et al., 2002), e.g. by deflection due to the horizontal component of the Coriolis force.

Comparing the temperature time series with those of waterflow speed (Fig. 2b) shows that the relatively warm SW seem to be related with the dominant 0.05-m s$^{-1}$ amplitude variations occurring with about 20-day periodicity. These relatively broadband (not shown) quasi-periodic waterflow-speed variations are considered to be related to the dynamically unstable meandering Northern boundary Current and associated mesoscale eddies (Crepon et al., 1982). Although the waterflow is generally strongest near the surface, it can still be measured at great depths. Spatial flow varies over only 50 m horizontally. Peak variations are associated in half the cases in their absolute value with SW (not shown). These flow differences represent smaller than sub-mesoscale vorticity and/or divergence. The SW in temperature also correlates in-phase with the relative acoustic amplitude (Fig. 2c), which reflects the amount of suspended particles and zooplankton, and by a delay of about 5 days with the near-local wind speed (Fig. 2d), as established from lag-correlation analysis. The wind-speed squared (wind-load) is observed to be 20% higher than the average value in winter, while 13% lower than the average value in summer.

This seasonal variation in wind-load is reflected in 124-m vertically, 45-line horizontally, and daily averaged deep-sea turbulence dissipation rates (Fig. 2e), which are about 25% higher than the year-long average value in winter and 14% lower than the average value in summer. The seasonal variation in wind-load also reflects that turbulence dissipation rates are generally larger under SW than under NH conditions, on average by a factor of 3.5 (Fig. 2e). NH are dominated by buoyancy-driven GH-fluxes through the seafloor that provide average turbulence dissipation rate of $\varepsilon_{GH}=1.2\times10^{-10}$ m$^2$ s$^{-3}$ in the area. SW are dominated by internal-wave breaking driven by sub-mesoscale and near-inertial motions.

To compare SW-turbulence values with those possibly generated by deep dense-water formation a proxy is used from observed nighttime near-surface cooling (Brainerd & Gregg, 1995). To match SW-turbulence values, deep dense water should reach the ~2500-m deep seafloor during a period of about 2.5 months once every 8 years, or during 9 days every year, which is not observed.

The observed average turbulence dissipation rate over the year of $2.4\pm0.2\times10^{-10}$ m$^2$ s$^{-3}$ from the tide-deprived deep Western Mediterranean site is a factor 3-10 larger than found in the tide-dominated open ocean well away from boundaries (de Lavergne et al., 2020; Yasuda et al., 2021). The standard deviation reflects the variation between the 45 lines. Environmental variations with time provide a larger spread of values and result in a larger apparent standard deviation for single-line values.

One of the aims of the mooring-array was to enable investigations of deep-sea turbulence via short movies (Fig. 3; movies in: van Haren & KM3NeT Collaboration, 2025a). Six examples are presented, three categorised as SW and three as NH. Several characteristics including mean waterflows are given in Table 1. In each movie, progress with time is visualised by a white line moving in a time-depth panel from line 24. This panel is above the cube of flashing T-sensors. For the cube, the fixed viewpoint is lifted up near the east-southeasterly corner. All T-sensors are visible in this view, but one needs to capture the perspective of 3D-interpretation in the 2D-projected cube. Waterflow measurements are only made at h=126 m, which is just above the top of the cube.

In Fig. 3a a common SW period is shown, with typical vertical temperature difference and turbulence dissipation rate (Table 1). The movie has a calm start for $t < 307.66$ with gentle short-'wave' temperature variations crossing the lines near their top. The smaller than one hour periodicity is too short for freely propagating internal waves, and the motions are in the turbulence range mainly. Relatively warm waters are seen to enter from above plunging about 100 m down and reaching the seafloor. They briefly generate relatively strong turbulence. The warm waters are replaced by a turbulent cold front near the seafloor coming from the northeast with trailing slow internal waves near the local buoyancy frequency and small-scale turbulence jitter. To capture motion-directions, repeated playing and rewinding of the movie is helpful.

In Fig. 3b characteristics of NH are shown with a one order of magnitude smaller temperature range than in Fig. 3a. Putting the deep-sea motions in perspective: hypothetical particles advected by this flow cross the cube in about 0.014 day, which is 2 s in movie time. Similar to the developments displayed in Fig 3a also here a calm start is observed, this time with a mid-depth turbulent cloud, which is replaced by a warming from above across the T-sensor range. In contrast to the developments shown in Fig. 3a, the cold-water replacement enters at mid-height. Above and below small-scale warming motions occur. Near the top, a slow wavy-like turbulence pattern is seen. Near the seafloor GH's horizontal scales are small and differences are seen per line. This is similar to the boiling in a pot that is uniformly heated from below, but for which the fluid responds with varying positions of small-scale bubble lines designating the upgoing motions (especially after day 356.9). The 9.5-m horizontal separation between lines is at least two orders of magnitude smaller than reported underground geological vents that feed the heating from below (Kunath et al., 2021). Despite the considerable mean waterflow near the top, most small-scale motions near the seafloor are found to be strictly vertical, and vertically suppressed presumably by weakly stratified waters above. Towards the end of this movie, GH flares up at the upper T-sensors and leaves turbulent clouds behind that are advected by the waterflow out of the cube.

The NH example shown in Fig. 3c displays more large-scale GH, which frequently varies in height and in intensity. Some heated water and turbulent clouds are advected with the mean flow across the cube, e.g., around day 495.73. Advection seems to occur at some distance, about h>30 m, from the seafloor. As in Fig. 3b, most convection-turbulence motions are found vertically varying with time, and between neighbouring lines.

In Fig 3d one of the most turbulent SW periods is seen, between days 606.9 and 607.1. It is one order of magnitude more turbulent than the average. Initially, warm water comes from above with fast waves of turbulent motions. The motions have quasi-periodicities that are tenfold larger than the maximum small-scale buoyancy period. They unlikely represent freely propagating internal waves. They are followed by fast moving turbulent clouds from the side mostly from the northeast, with apparent fast half turn and slanting from above/sideways of warm waters into the cube around day 606.97. Near the seafloor cooler water comes from the west-northwest. The second half of the movie shows calmer internal waves, with the largest one having a periodicity commensurate with maximum small-scale buoyancy frequency, but remaining turbulently jittering with higher frequent quasi-periodic overturn motions. In the blue, a two-arm front passes over the seafloor in south-southwesterly direction between days 607.12 and 607.15.

In Fig. 3e interfacial-wave turbulence of various scales and in various propagation directions is demonstrated. The position of the interface of relatively strong stratification in a thin layer varies with time about 50 m vertically. The horizontal extension of the array is large enough to resolve the amplitudes of the motion. Across interfaces, the 2-m-small-scale buoyancy frequency reaches a maximum of $N_s = 6f$. Thus, freely propagating internal waves have shortest measurable periods of about 3 hours (1.3 times the bar in Fig. 3b). No information is available about possible smaller-scale buoyancy frequencies, because of the 2-m interval between T-sensors. Faster fluctuating turbulent motions are pumping the interface up and down at about 10-m scales in counter-phase between neighbouring lines, best visible in the cyan-coloured layer, e.g. between 468.4 and 468.45. Just after the start of the movie at day 468.35 a cyan cloud is transported to the west, thereby forcing the interface up generating a wavy motion (best seen in the time-depth plot for line 24). The apparent up- and down-going standing wave motions represent transitions to turbulent overturns under sufficient shear for growth of instabilities (Bogdanov et al., 2023). These transitions are reminiscent of the process of parametric instability in which primary, freely propagating waves disintegrate into mode-2 oscillations of much shorter length scale. Mode-2 oscillations are distinguished by counter-moving isotherms. Such a transition process has been shown theoretically and in a laboratory model (Davis and Acrivos, 1967). In the present deep-sea data, the scale of the standing waves exceeds 10 m, as neighbouring lines consistently show out-of-phase motions in time in all directions, with widening and shrinking in colour bands in the vertical. This can better be seen in the enlargement of reference-line 24 and its three neighbours, e.g. in the purple and orange ellipses around a single large overturn of Fig. 4. This mechanism of turbulence generation is observed at many different time and length scales. The mean turbulence dissipation rate of such interior internal-wave breaking amounts to $3\pm1\times10^{-11}$ $m^2$ $s^{-3}$, which is equivalent to open-ocean turbulence measurements (Yasuda et al., 2021) and one order of magnitude smaller than average slantwise warming. During the second half of the movie, GH becomes apparent near the seafloor, but is not well visible with the used colour-coding.

In Fig. 3f NH is presented that is not clearly dominated by GH. Via warming from above, regular gentle turbulence is spread over the lower 100 m above the seafloor. The movie seems in slow-motion compared to the previous movies, but it has the same speed. The main variation in temperature has a period of about 0.35 day, which is half the inertial period. Turbulent overturns of different sizes are visible, clouds of warmer water are deposited at the seafloor, move partially up, and disperse out of the cube. While in the time-depth panel on the top slanted spurs of relatively warm water are visible, these are seen in the movie to go up or down, as well as become advected out of the cube. They represent local turbulent overturns with scales O(10) m.

## 3. Discussion and Conclusions

The large-ring mooring provides observational insight in 3D and time development of deep-sea turbulence in an area where stratification is generally weak. Besides improved statistics of spectral information, to be investigated in near-future, and of turbulence dissipation rate, the large-ring mooring covers a reasonable range of sizes to distinguish internal-wave and geothermal-heat induced turbulence development as inferred from half-day real-time-compressed movies. This information cannot be obtained by common 1D oceanographic observations, or in 2D. Thus, the large-ring mooring provides sufficiently new insights, potentially also at other ocean areas in the future. The 9.5-m horizontal and 2-m vertical scales seem adequate for the purpose. However, for resolution of entire internal-tidal excursion lengths of about one kilometre in the deep sea a larger set-up is required. To turn the mooring in a cabled observatory requires a new design and construction of T-sensors.

Unfortunately, the 20-month-long records do not register the sinking of unstable cool waters, which may be rare at the mooring site close to the continental slope. Instead of sinking of cool water, the T-sensor data show a dominance of warm-water pushes from above and sideways and, when stratification is negligible, GH-induced convection-turbulence. GH can only be detected when internal-wave action from above is low. GH's water response is found to vary strongly in height, from O(10) to O(100) m, with equivalent horizontal sizes so that small-scale response is different even between neighbouring lines. Waters heated from below are advected horizontally when they reach at least 30 m above seafloor. GH's mean dissipation rate (van Haren, 2025a) matches the local average mean heat flux of 0.11 W $m^{-2}$ determined from geophysics measurements (Pasquale et al., 1996) and which are considered important for deep-sea circulation (Adcroft et al., 2001; Park et al., 2013; Ferron et al., 2017). Besides GH, three more processes of turbulence generation are distinguished.

The case of warm-water input from above conceptualises as a stably stratified condition, which is previously thought to suppress turbulence. However, it is inferred that such a case can provide about 3.5 times larger average turbulence in the deep sea than GH, which contrasts with previous findings (Feron et al., 2017) for the entire Northwestern Mediterranean and is closer to their Ligurian-Sea data . The large turbulence is not representing deep dense-water formation, because it is observed year-round and because associated intensification of waterflow speed is not observed. As warming occurs sufficiently often, about 40% of time versus 60% for NH conditions, it dominates the turbulence generation at the mooring site. It is about one order of magnitude more energetic than average open-ocean turbulence away from boundaries as inferred from low-noise-profiler data (e.g., Yasuda et al., 2021). Occasional tenfold more energetic turbulence than average is observed generated by internal-wave motions with amplitudes exceeding the 124-m range of observations, and resulting in turbulent clouds that are advected through the mooring-array. Thus SW conditions generate considerably more turbulent mixing in the deep sea over a flat seafloor than hitherto thought, which is important for organisms living there.

Processes of relatively weak turbulence ten times smaller than the SW-average are found in two different cases. In the case of near-homogeneous waters, turbulence is seen to deposit warm-water blobs at the seafloor that subsequently rise again from the seafloor as in convection. In the case of relatively strong, compared to the large-scale average, thin-layer stratification, propagating primary internal waves are seen to parametrically generate short-scale mode-2 standing wave motions of 10-m horizontal scales that break locally. Such interfacial internal-wave instabilities were observed to occur in laboratory models and shallow seas (Davis and Acrivos, 1967; Bogdanov et al., 2023). However, they occur also in the generally weakly-stratified deep sea where interfaces of local strong stratification are expected to be rare. Considering their observed mean turbulence value, this mechanism may explain the weak open-ocean turbulence away from boundaries.

A coupling with variations on a seasonal scale is suggested between SW-conditions including associated turbulence and wind-load. The 2500-m deep-sea response trails the surface by about one week. The pathway of energy transfer from an atmospheric disturbance passage is via formation of mesoscale, sub-mesoscale eddies and internal wave motions. In the area, anticyclonic eddies live longest, typically one year (Testor & Gascard, 2006). Such eddies trap near-inertial motions (Kunze, 1985). These lead to transfer of warmer water to approach the seafloor, breaking waves and subsequent turbulence generation, most likely via slantwise convection (Marshall and Schott, 1999; Straneo et al., 2002). As convection occurs in up- and down-going plumes, the slantwise deflection in the direction of the Earth rotation vector results in local apparent minimum large-scale buoyancy frequency of $N_{min} \approx 2f$ (van Haren, 2008), as observed. A transfer from 1000-km atmosphere scale, to near-surface 100 and 10-km eddies, and sea-interior 1-km inertial-wave scale leads via strong nonlinear interactions to deep-sea 100-m and smaller-scale internal waves and, eventually, convection-turbulence overturning. The observed turbulence mechanisms locally mix suspended matter, oxygen and replenishes food for different life forms. This seems to be completely sufficient to maintain the food supply, as for this year no local deep dense water formation had been observed.

## Appendix: Authors of the KM3NeT Collaboration

H. van Haren[ar,*], O. Adriani[b,a], A. Albert[c,bd], A.R. Alhebsi[d], S. Alshalloudi[d], M. Alshamsi[e], S. Alves Garre[f], A. Ambrosone[h,g], F. Ameli[i], M. Andre[j], L. Aphecetche[k], M. Ardid[l], S. Ardid[l], J. Aublin[m], F. Badaracco[o,n], L. Bailly-Salins[p], B. Baret[m], A. Bariego-Quintana[f], Y. Becherini[m], M. Bendahman[g], F. Benfenati Gualandi[r,q], M. Benhassi[g,s], D.M. Benoit[t], Z. Beňušová[v,u], E. Berbee[w], E. Berti[b], V. Bertin[e], P. Betti[b], S. Biagi[x], M. Boettcher[y], D. Bonanno[x], S. Bottai[b], A.B. Bouasla[be], J. Boumaaza[z], M. Bouta[e], M. Bouwhuis[w], C. Bozza[aa,g], R.M. Bozza[h,g], H.Brânzaş[ab], F. Bretaudeau[k], M. Breuhaus[e], R. Bruijn[ac,w], J. Brunner[e], R. Bruno[ad], E. Buis[ae,w], R. Buompane[s,g], J. Busto[e], B. Caiffi[o] D. Calvo[f], A. Capone[i,af], F. Carenini[r,q], V. Carretero[ac,w], T. Cartraud[m], P. Castaldi[ag,q], V. Cecchini[f], S. Celli[i,af], L. Cerisy[e], M. Chabab[ah], A. Chen[ai], S. Cherubini[aj,x], T. Chiarusi[q], M. Circella[ak], R. Clark[al], R. Cocimano[x], J.A.B. Coelho[m], A. Coleiro[m], A. Condorelli[m], R. Coniglione[x], P. Coyle[e], A. Creusot[m], G. Cuttone[x], R. Dallier[k], A. De Benedittis[s,g], G. De Wasseige[al], V. Decoene[k], P. Deguire[e], I. Del Rosso[r,q], L.S. Di Mauro[x], I. Di Palma[i,af], A.F. Díaz[am], D. Diego-Tortosa[x], C. Distefano[x], A. Domi[an], C. Donzaud[m], D. Dornic[e], E. Drakopoulou[ao], D. Drouhin[c,bd], J.-G. Ducoin[e], P. Duverne[m], R. Dvornický[v], T. Eberl[an], E. Eckerová[v,u], A. Eddymaoui[z], T. van Eeden[w], M. Eff[m], D. van Eijk[w], I. El Bojaddaini[ap], S. El Hedri[m], S. El Mentawi[e], A. Enzenhöfer[e], G. Ferrara[aj,x], M. D. Filipović[aq], F. Filippini[q], D. Franciotti[x], L.A. Fusco[aa,g], T. Gal[an], J. García Méndez[l], A. Garcia Soto[f], C. Gatius Oliver[w], N. Geißelbrecht[an], E. Genton[al], H. Ghaddari[ap], L. Gialanella[s,g], B.K. Gibson[t], E. Giorgio[x], I. Goos[m], P. Goswami[m], S.R. Gozzini[f], R. Gracia[an], B. Guillon[p], C. Haack[an], A. Heijboer[w], L. Hennig[an], J. J. Hernández-Rey[f], A. Idrissi[x], W. Idrissi Ibnsalih[g], G. Illuminati[q], R. Jaimes[f], O. Janik[an], D. Joly[e], M. de Jong[as,w], P. de Jong[ac,w], B.J. Jung[w], P. Kalaczyński[bf,at], J. Keegans[t], V. Kikvadze[au], G. Kistauri[av,au], C. Kopper[an], A. Kouchner[aw,m], Y.Y. Kovalev[ax], L. Krupa[u], V. Kueviakoe[w], V. Kulikovskiy[o], R. Kvatadze[av], M. Labalme[p], R. Lahmann[an], M. Lamoureux[al], G. Larosa[x], C. Lastoria[p], J. Lazar[al], A. Lazo[f], G. Lehaut[p], V. Lemaître[al], E. Leonora[ad], N. Lessing[f], G. Levi[r,q], M. Lindsey Clark[m], F. Longhitano[ad], S. Madarapu[f], F. Magnani[e], L. Malerba[o,n], F. Mamedov[u], A. Manfreda[g], A. Manousakis[ay], M. Marconi[n,o], A. Margiotta[r,q], A. Marinelli[h,g], C. Markou[ao], L. Martin[k], M. Mastrodicasa[af,i], S. Mastroianni[g], J. Mauro[al], K.C.K. Mehta[at], G. Miele[h,g], P. Migliozzi[g], E. Migneco[x], M.L. Mitsou[s,g], C.M. Mollo[g], L. Morales-Gallegos[s,g], N. Mori[b], A. Moussa[ap], I. Mozun Mateo[p], R. Muller[q], M.R. Musone[s,g], M. Musumeci[x], S. Navas[az], A. Nayerhoda[ak], C.A. Nicolau[i], B. Nkosi[ai], B. Ó Fearraigh[o], V. Oliviero[h,g], A. Orlando[x], E. Oukacha[m], L. Pacini[b], D. Paesani[x], J. Palacios González[f], G. Papalashvili[ak,au], P. Papini[b], V. Parisi[n,o], A. Parmar[p], C. Pastore[ak], A.M. Păun[ab], G.E. Păvălaş[ab], S. Peña Martínez[m], M. Perrin-Terrin[e], V. Pestel[p], M. Petropavlova[u,bg], P. Piattelli[x], A. Plavin[ax,bh], C. Poirè[aa,g], V. Popa[†ab], T. Pradier[c], J. Prado[f], S. Pulvirenti[x], C.A. Quiroz-Rangel[l], N. Randazzo[ad], A. Ratnani[ba], S. Razzaque[bb], I.C. Rea[g], D. Real[f], G. Riccobene[x], J. Robinson[y], A. Romanov[n,o,p], E. Ros[ax], A. Šaina[f], F. Salesa Greus[f], D.F.E. Samtleben[as,w], A. Sánchez Losa[f], S. Sanfilippo[x], M. Sanguineti[n,o], D. Santonocito[x], P. Sapienza[x], M. Scaringella[b], M. Scarnera[al,m], J. Schnabel[an], J. Schumann[an], J. Seneca[w], P.A. Sevle Myhr[al], I. Sgura[ak], R. Shanidze[au], Chengyu Shao[bi,e], A. Sharma[m], Y. Shitov[u], F. Šimkovic[v], A. Simonelli[g], A. Sinopoulou[ad], B. Spisso[g], M. Spurio[r,q], O. Starodubtsev[b], D. Stavropoulos[ao], I. Štekl[u], D. Stocco[k], M. Taiuti[n,o], G. Takadze[au], Y. Tayalati[z,ba], H. Thiersen[y], S. Thoudam[d], I. Tosta e Melo[ad,aj], B. Trocmé[m], V. Tsourapis[ao], E. Tzamariudaki[ao], A. Ukleja[at], A. Vacheret[p], V. Valsecchi[x], V. Van Elewyck[aw,m], G. Vannoye[n,o], E. Vannuccini[b], G. Vasileiadis[bc], F. Vazquez de Sola[w], A. Veutro[i,af], S. Viola[x], D. Vivolo[s,g], A. van Vliet[d], E. de Wolf[ac,w], I. Lhenry-Yvon[m], S. Zavatarelli[o], D. Zito[x], J. D. Zornoza[f], J. Zúñiga[f]

*Corresponding author: Hans van Haren, hans.van.haren@nioz.nl; km3net-pc@km3net.de

[a] *Università di Firenze, Dipartimento di Fisica e Astronomia, via Sansone 1, Sesto Fiorentino, 50019 Italy*

[b] *INFN, Sezione di Firenze, via Sansone 1, Sesto Fiorentino, 50019 Italy*

[c] *Université de Strasbourg, CNRS, IPHC UMR 7178, F-67000 Strasbourg, France*

[d] *Khalifa University of Science and Technology, Department of Physics, PO Box 127788, Abu Dhabi, United Arab Emirates*

[e] *Aix Marseille Univ, CNRS/IN2P3, CPPM, Marseille, France 72*

[f] *IFIC - Instituto de Física Corpuscular (CSIC - Universitat de València), c/Catedrático José Beltrán, 2, 46980 Paterna, Valencia, Spain*

† Deceased

[g]INFN, Sezione di Napoli, Complesso Universitario di Monte S. Angelo, Via Cintia ed. G, Napoli, 80126 Italy
[h]Università di Napoli "Federico II", Dip. Scienze Fisiche "E. Pancini", Complesso Universitario di Monte S. Angelo, Via Cintia ed. G, Napoli, 80126 Italy
[i]INFN, Sezione di Roma, Piazzale Aldo Moro, 2 - c/o Dipartimento di Fisica, Edificio, G.Marconi, Roma, 00185 Italy
[j]Universitat Politècnica de Catalunya, Laboratori d'Aplicacions Bioacústiques, Centre Tecnològic de Vilanova i la Geltrú, Avda. Rambla Exposició, s/n, Vilanova i la Geltrú, 08800 Spain
[k]Subatech, IMT Atlantique, IN2P3-CNRS, Nantes Université, 4 rue Alfred Kastler - La Chantrerie, Nantes, BP 20722 44307 France
[l]Universitat Politècnica de València, Instituto de Investigación para la Gestión Integrada de las Zonas Costeras, C/ Paranimf, 1, Gandia, 46730 Spain
[m]Université Paris Cité, CNRS, Astroparticule et Cosmologie, F-75013 Paris, France
[n]Università di Genova, Via Dodecaneso 33, Genova, 16146 Italy
[o]INFN, Sezione di Genova, Via Dodecaneso 33, Genova, 16146 Italy
[p]LPC CAEN, Normandie Univ, ENSICAEN, UNICAEN, CNRS/IN2P3, 6 boulevard Maréchal Juin, Caen, 14050 France
[q]INFN, Sezione di Bologna, v.le C. Berti-Pichat, 6/2, Bologna, 40127 Italy
[r]Università di Bologna, Dipartimento di Fisica e Astronomia, v.le C. Berti-Pichat, 6/2, Bologna, 40127 Italy
[s]Università degli Studi della Campania "Luigi Vanvitelli", Dipartimento di Matematica e Fisica, viale Lincoln 5, Caserta, 81100 Italy
[t]E. A. Milne Centre for Astrophysics, University of Hull, Hull, HU6 7RX, United Kingdom
[u]Czech Technical University in Prague, Institute of Experimental and Applied Physics, Husova 240/5, Prague, 110 00 Czech Republic
[v]Comenius University in Bratislava, Department of Nuclear Physics and Biophysics, Mlynska dolina F1, Bratislava, 842 48 Slovak Republic
[w]Nikhef, National Institute for Subatomic Physics, PO Box 41882, Amsterdam, 1009 DB Netherlands
[x]INFN, Laboratori Nazionali del Sud, (LNS) Via S. Sofia 62, Catania, 95123 Italy
[y]North-West University, Centre for Space Research, Private Bag X6001, Potchefstroom, 2520 South Africa
[z]University Mohammed V in Rabat, Faculty of Sciences, 4 av. Ibn Battouta, B.P. 1014, R.P. 10000 Rabat, Morocco
[aa]Università di Salerno e INFN Gruppo Collegato di Salerno, Dipartimento di Fisica, Via Giovanni Paolo II 132, Fisciano, 84084 Italy
[ab]Institute of Space Science - INFLPR Subsidiary, 409 Atomistilor Street, Magurele, Ilfov, 077125 Romania
[ac]University of Amsterdam, Institute of Physics/IHEF, PO Box 94216, Amsterdam, 1090 GE Netherlands
[ad]INFN, Sezione di Catania, (INFN-CT) Via Santa Sofia 64, Catania, 95123 Italy
[ae]TNO, Technical Sciences, PO Box 155, Delft, 2600 AD Netherlands
[af]Università La Sapienza, Dipartimento di Fisica, Piazzale Aldo Moro 2, Roma, 00185 Italy
[ag]Università di Bologna, Dipartimento di Ingegneria dell'Energia Elettrica e dell'Informazione "Guglielmo Marconi", Via dell'Università 50, Cesena, 47521 Italia
[ah]Cadi Ayyad University, Physics Department, Faculty of Science Semlalia, Av. My Abdellah, P.O.B. 2390, Marrakech, 40000 Morocco
[ai]University of the Witwatersrand, School of Physics, Private Bag 3, Johannesburg, Wits 2050 South Africa
[aj]Università di Catania, Dipartimento di Fisica e Astronomia "Ettore Majorana", (INFN-CT) Via Santa Sofia 64, Catania, 95123 Italy
[ak]INFN, Sezione di Bari, via Orabona, 4, Bari, 70125 Italy
[al]UCLouvain, Centre for Cosmology, Particle Physics and Phenomenology, Chemin du Cyclotron, 2, Louvain-la-Neuve, 1348 Belgium

[am]*University of Granada, Department of Computer Engineering, Automation and Robotics / CITIC, 18071 Granada, Spain*
[an]*Friedrich-Alexander-Universität Erlangen-Nürnberg (FAU), Erlangen Centre for Astroparticle Physics, Nikolaus-Fiebiger-Straße 2, 91058 Erlangen, Germany*
[ao]*NCSR Demokritos, Institute of Nuclear and Particle Physics, Ag. Paraskevi Attikis, Athens, 15310 Greece*
[ap]*University Mohammed I, Faculty of Sciences, BV Mohammed VI, B.P. 717, R.P. 60000 Oujda, Morocco*
[aq]*Western Sydney University, School of Science, Locked Bag 1797, Penrith, NSW 2751 Australia*
[ar]*NIOZ (Royal Netherlands Institute for Sea Research), PO Box 59, Den Burg, Texel, 1790 AB, the Netherlands*
[as]*Leiden University, Leiden Institute of Physics, PO Box 9504, Leiden, 2300 RA Netherlands*
[at]*AGH University of Krakow, Al. Mickiewicza 30, 30-059 Krakow, Poland*
[au]*Tbilisi State University, Department of Physics, 3, Chavchavadze Ave., Tbilisi, 0179 Georgia*
[av]*The University of Georgia, Institute of Physics, Kostava str. 77, Tbilisi, 0171 Georgia*
[aw]*Institut Universitaire de France, 1 rue Descartes, Paris, 75005 France*
[ax]*Max-Planck-Institut f ̈ur Radioastronomie, Auf dem Hügel 69, 53121 Bonn, Germany*
[ay]*University of Sharjah, Sharjah Academy for Astronomy, Space Sciences, and Technology, University Campus - POB 27272, Sharjah, - United Arab Emirates*
[az]*University of Granada, Dpto. de Física Teórica y del Cosmos & C.A.F.P.E., 18071 Granada, Spain*
[ba]*School of Applied and Engineering Physics, Mohammed VI Polytechnic University, Ben Guerir, 43150, Morocco*
[bb]*University of Johannesburg, Department Physics, PO Box 524, Auckland Park, 2006 South Africa*
[bc]*Laboratoire Univers et Particules de Montpellier, Place Eugène Bataillon - CC 72, Montpellier Cédex 05, 34095 France*
[bd]*Université de Haute Alsace, rue des Frères Lumière, 68093 Mulhouse Cedex, France*
[be]*Université Badji Mokhtar, Département de Physique, Faculté des Sciences, Laboratoire de Physique des Rayonnements, B. P. 12, Annaba, 23000 Algeria*
[bf]*AstroCeNT, Nicolaus Copernicus Astronomical Center, Polish Academy of Sciences, Rektorska 4, Warsaw, 00-614 Poland*
[bg]*Charles University, Faculty of Mathematics and Physics, Ovocný trh 5, Prague, 116 36 Czech Republic*
[bh]*Harvard University, Black Hole Initiative, 20 Garden Street, Cambridge, MA 02138 USA*
[bi]*School of Physics and Astronomy, Sun Yat-sen University, Zhuhai, China*

**Conflict of Interest**
The authors declare no conflict of interest relevant to this study.

**Open Research**
Current meter and CTD data are available from van Haren (2025b). Data to figures 2-4 are available from van Haren & KM3NeT Collaboration (2025b). Movies to Fig. 3 are available in: van Haren & KM3NeT Collaboration (2025a). Atmospheric data are retrieved from https://content.meteoblue.com/en/business-solutions/weather-apis/dataset-api.

**Acknowledgements** We thank captains and crews of R/V Pelagia for the very pleasant cooperation. We also thank the team of ROV Holland I for the well-performed underwater mission to recover the instrumentation of the large ring. NIOZ colleagues notably from NMF department are thanked for their indispensable contributions during the long preparatory and construction phases to make the unique sea-operation successful. The authors acknowledge the financial support of: KM3NeT-INFRADEV2 project, funded by the European Union Horizon Europe Research and Innovation Programme under grant agreement No 101079679; Funds for Scientific Research (FRS-FNRS), Francqui foundation, BAEF foundation. Czech Science Foundation (GAČR 24-12702S); Agnce Nationale de la Recherche (contract ANR-15-CE31-0020), Centre National de la Recherche Scientifique (CNRS), Commission Européenne (FEDER fund and Marie Curie Program), LabEx UnivEarthS (ANR-10-LABX-0023 and ANR-18-IDEX-0001), Paris Île-de-France Region, Normandy Region (Alpha, Blue-waves and Neptune), France, The Provence-Alpes-Côte d'Azur Delegation for Research and Innovation (DRARI), the Provence-Alpes-Côte d'Azur region, the Bouches-du-Rhône Departmental Council, the Metropolis of Aix-Marseille Provence and the City of Marseille through the CPER 2021-2027 NEUMED project, The CNRS Institut National de Physique Nucléaire et de Physique des Particules (IN2P3); Shota Rustaveli National Science Foundation of Georgia (SRNSFG, FR-22-13708), Georgia; This work is part of the MuSES project which has received funding from the European Research Council (ERC) under the European Union's Horizon 2020 Research and Innovation Programme (grant agreement No 101142396). The General Secretariat of Research and Innovation (GSRI), Greece; Istituto Nazionale di Fisica Nucleare (INFN) and Ministero dell'Università e della Ricerca (MUR), through PRIN 2022 program (Grant PANTHEON 2022E2J4RK, Next Generation EU) and PON R&I program (Avviso n. 424 del 28 febbraio 2018, Progetto PACK-PIR01 00021), Italy; IDMAR project Po-Fesr Sicilian Region az. 1.5.1; A. De Benedittis, W. Idrissi Ibnsalih, M. Bendahman, A. Nayerhoda, G. Papalashvili, I. C. Rea, A. Simonelli have been supported by the Italian Ministero dell'Università e della Ricerca (MUR), Progetto CIR01 00021 (Avviso n. 2595 del 24 dicembre 2019); KM3NeT4RR MUR Project National Recovery and Resilience Plan (NRRP), Mission 4 Component 2 Investment 3.1, Funded by the European Union – NextGenerationEU,CUP I57G21000040001, Concession Decree MUR No. n. Prot. 123 del 21/06/2022; Ministry of Higher Education, Scientific Research and Innovation, Morocco, and the Arab Fund for Economic and Social Development, Kuwait; Nederlandse organisatie voor Wetenschappelijk Onderzoek (NWO), the Netherlands; The grant "AstroCeNT: Particle Astrophysics Science and Technology Centre", carried out within the International Research Agendas programme of the Foundation for Polish Science financed by the European Union under the European Regional Development Fund; The program: "Excellence initiative-research university" for the AGH University in Krakow; The ARTIQ project: UMO-2021/01/2/ST6/00004 and ARTIQ/0004/2021; Ministry of Research, Innovation and Digitalisation, Romania; Slovak Research and Development Agency under Contract No. APVV-22-0413; Ministry of Education, Research, Development and Youth of the Slovak Republic; MCIN for PID2021-124591NB-C41, -C42, -C43 and PDC2023-145913-I00 funded by MCIN/AEI/10.13039/501100011033 and by "ERDF A way of making Europe", for ASFAE/2022/014 and ASFAE/2022/023 with funding from the EU NextGenerationEU (PRTR-C17.I01) and Generalitat Valenciana, for Grant AST22 6.2 with funding from Consejería de Universidad, Investigación e Innovación and Gobierno de España and European Union - NextGenerationEU, for CSIC-INFRA23013 and for CNS2023-144099, Generalitat Valenciana for CIDEGENT/2020/049, CIDEGENT/2021/23, CIDEIG/2023/20, ESGENT2024/24, CIPROM/2023/51, GRISOLIAP/2021/192 and INNVA1/2024/110 (IVACE+i), Spain; Khalifa University internal grants (ESIG-2023-008, RIG-2023-070 and RIG-2024-047), United Arab Emirates; The European Union's Horizon 2020 Research and Innovation Programme (ChETEC-INFRA - Project no. 101008324). Views and opinions expressed are those of the author(s) only and do not necessarily reflect those of the European Union or the European Research Council. Neither the European Union nor the granting authority can be held responsible for them.

**Table 1.** General characteristics are given for half-day periods of Fig. 3, as indicated by their panel and start day. The 45-line, 124-m, and half-day mean turbulence dissipation rate values ε are given besides the full temperature range ΔΘ. Half-day mean waterflow speed |U| and direction (oceanographic convention) are provided from measurements at h = 126 m above seafloor. Two environmental conditions are defined: SW = stratified-water, NH = near-homogenous, and may include geothermal heat GH. The categories are distinguished by the equivalent of 124-m, half-day mean N = 0.65f.

| *Fig. 3* | *Start day* | *cond.* | *ε (m² s⁻³)* | *ΔΘ (°C)* | *\|U\| (m s⁻¹)* | *Dir.* |
|---|---|---|---|---|---|---|
| a | 307.6 | SW | $3.8\pm0.6\times10^{-10}$ | 0.003 | 0.03 | NNW |
| b | 356.55 | NH | $1.4\pm0.2\times10^{-10}$ | 0.00025 | 0.05 | ENE |
| c | 495.55 | NH | $1.1\pm0.2\times10^{-10}$ | 0.00025 | 0.05 | WNW |
| d | 606.8 | SW | $4.0\pm0.3\times10^{-9}$ | 0.0035 | 0.005 | SSE |
| e | 468.3 | SW | $1.0\pm0.1\times10^{-10}$ | 0.003 | 0.01 | SE |
| f | 647.96 | NH | $9\pm1\times10^{-11}$ | 0.0004 | 0.03 | ESE |

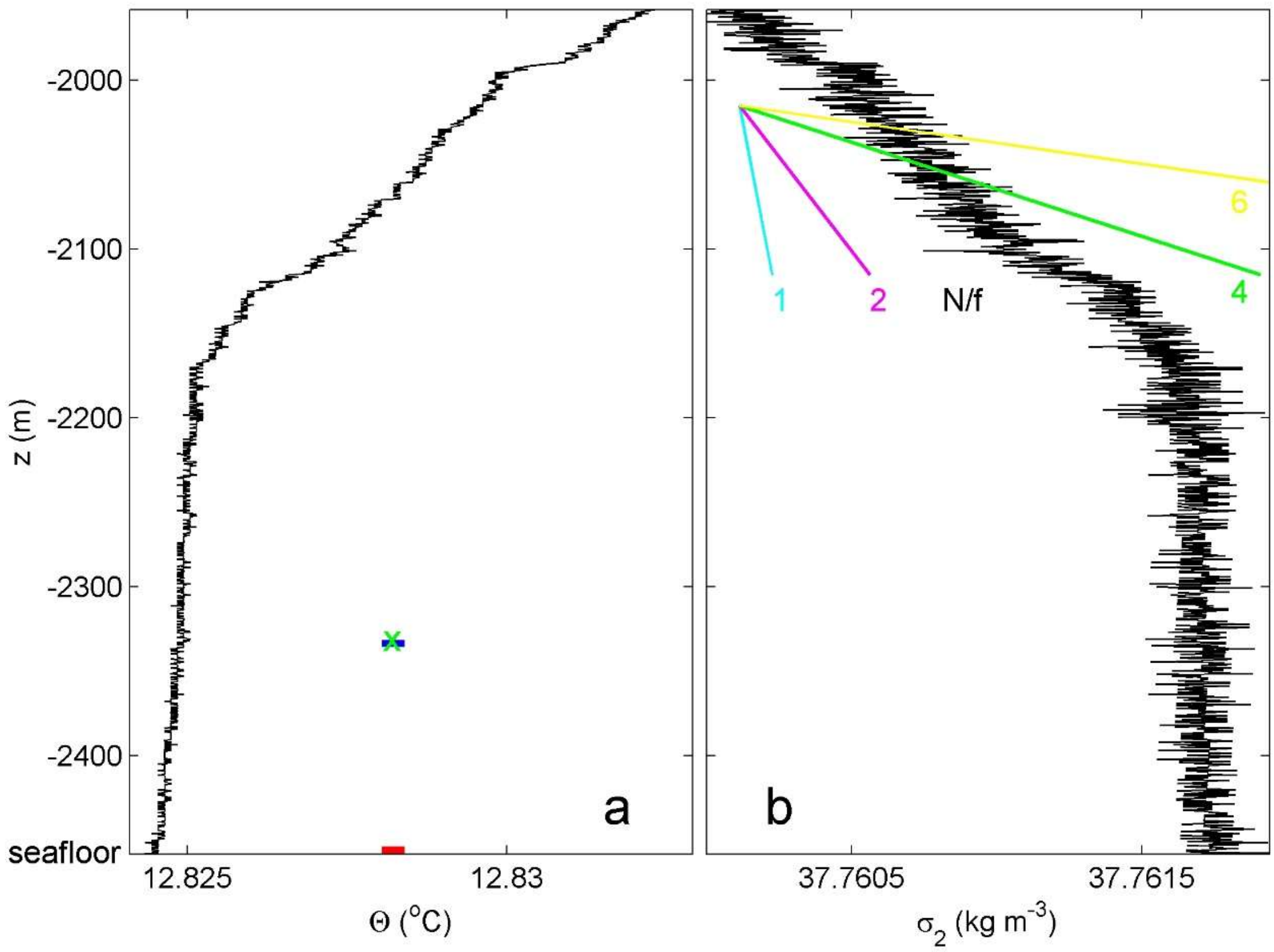

**Fig. 1.** Lower 500 m of single shipborne Conductivity Temperature Depth profile measured within height h = 0.5 m above seafloor, about 1 km from the large-ring mooring site during the deployment cruise in 2020. Small colour bars in panel a. indicate heights of uppermost (blue) and lowest (red) T-sensors in a mooring line, and the green cross the height of current meters. (a) Conservative Temperature (IOC et al., 2010). (b) Density anomaly −1000 kg $m^{-3}$ referenced to a pressure level of $2\times10^{7}$ N $m^{-2}$. The colour lines refer to slopes of different levels of 100-m vertical density stratification, and indicate the ratio of 'N' buoyancy over 'f' inertial frequency.

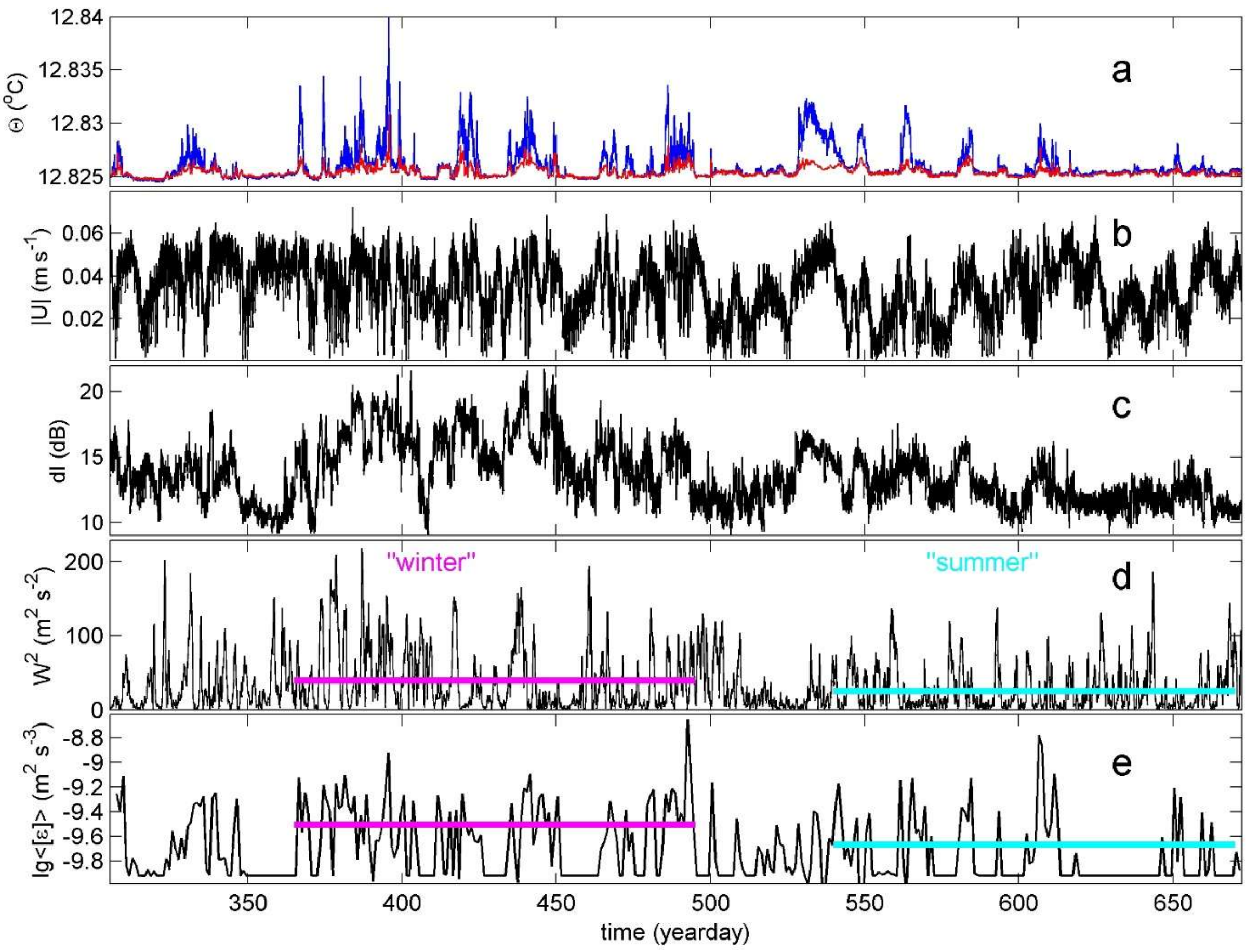

**Fig. 2.** Yearlong time series, mainly from data measured at the large-ring mooring. Time in days since beginning of 2020. (a) Detrended Conservative Temperature from lowest (red; h = 1 m above seafloor) and uppermost (blue; h = 125 m) T-sensors of line 24, sub-sampled at once per 10 s. (b) Hourly averaged waterflow speed at h = 126 m. (c) Relative acoustic amplitude measured at h = 126 m. (d) Wind-load measured at island-station Porquerolles, about 20 km north of the mooring site. The magenta and cyan bars indicate average values over 130 days. (e) Logarithm of daily, 124-m vertically, and 45-line averages of turbulence dissipation rate inferred from T-sensor data. During periods of near-homogeneous conditions, values are fixed at the mean value due to dominant geothermal heat flux.

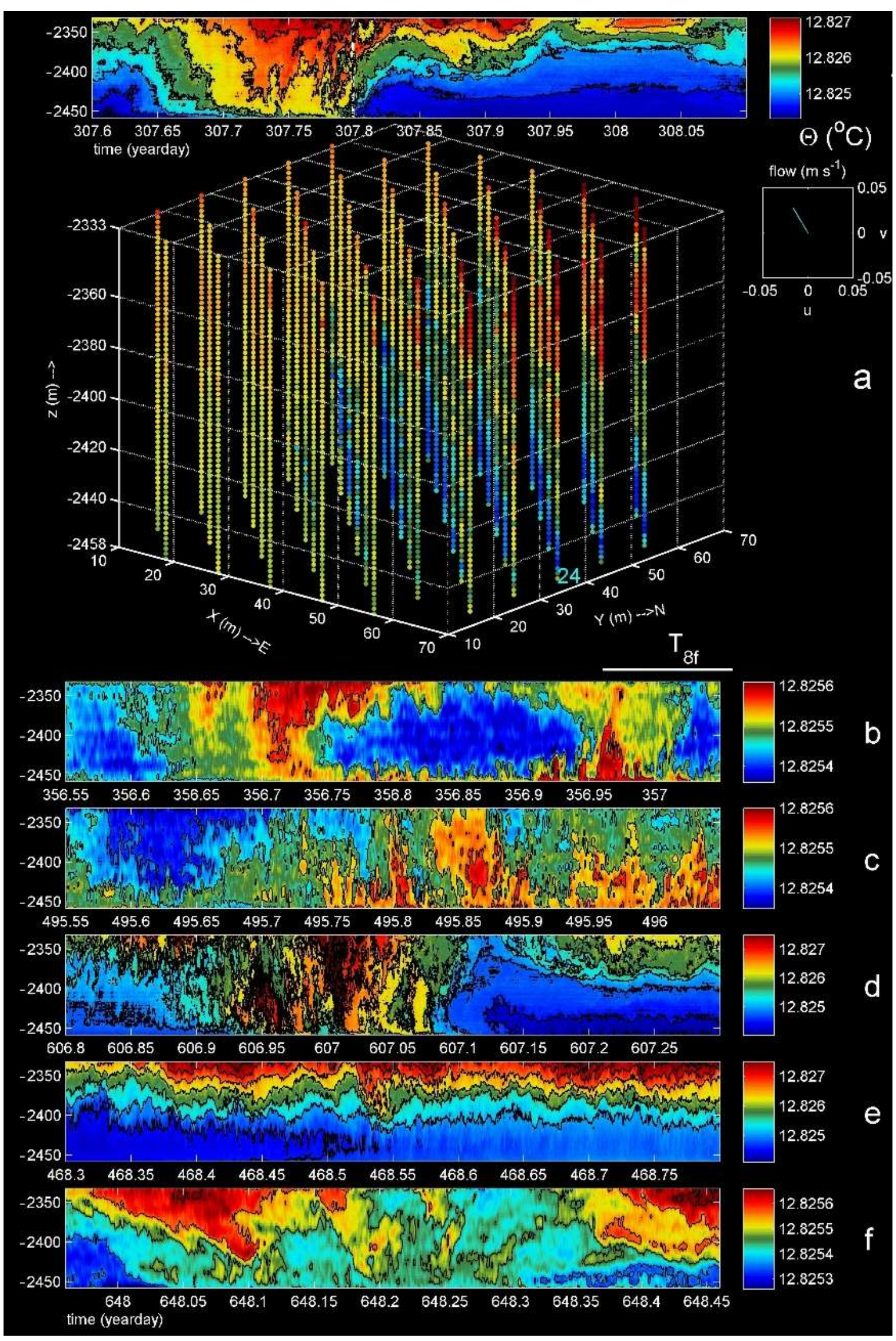

**Fig. 3.** Six quasi-3D movies from about 2800 T-sensors in nearly 0.5 hm$^3$ mooring-array. As indicated in a. only, each sensor is represented by a small filled circle, of which the colour indicates a temperature in the scale above. In the movies, above the cube, which is vertically depressed by about a factor of 2, a white time-line progresses in a half-day/124-m time/depth image from line 24 on the east side of the cube. The 72-s movies are accelerated by a factor of 600 with respect to real-time. Environmental characteristics are given in Table 1. (a) Full image of a movie. Example of stratified-water (SW) conditions. Upper-panel black contours are drawn at 0.0004°C intervals. (b) Time-depth image only of example of near-homogeneous (NH) conditions including mostly small suppressed geothermal-heating (GH) plumes. Contour interval is 0.0001°C. The white bar on top indicates the period of the overall maximum 2-m small-scale buoyancy period. (c) As b., but for NH including large GH plumes. Contour interval is 0.00005°C. (d) As b., but for strongly turbulent SW. Contour interval is 0.0004°C. (e) As b., but for SW interfacial internal-wave turbulent overturns above suppressed GH. Contour interval is 0.0005°C. (f) As b., but for double-inertial frequency waves and warm-water deposits under NH. Contour interval is 0.0001°C.

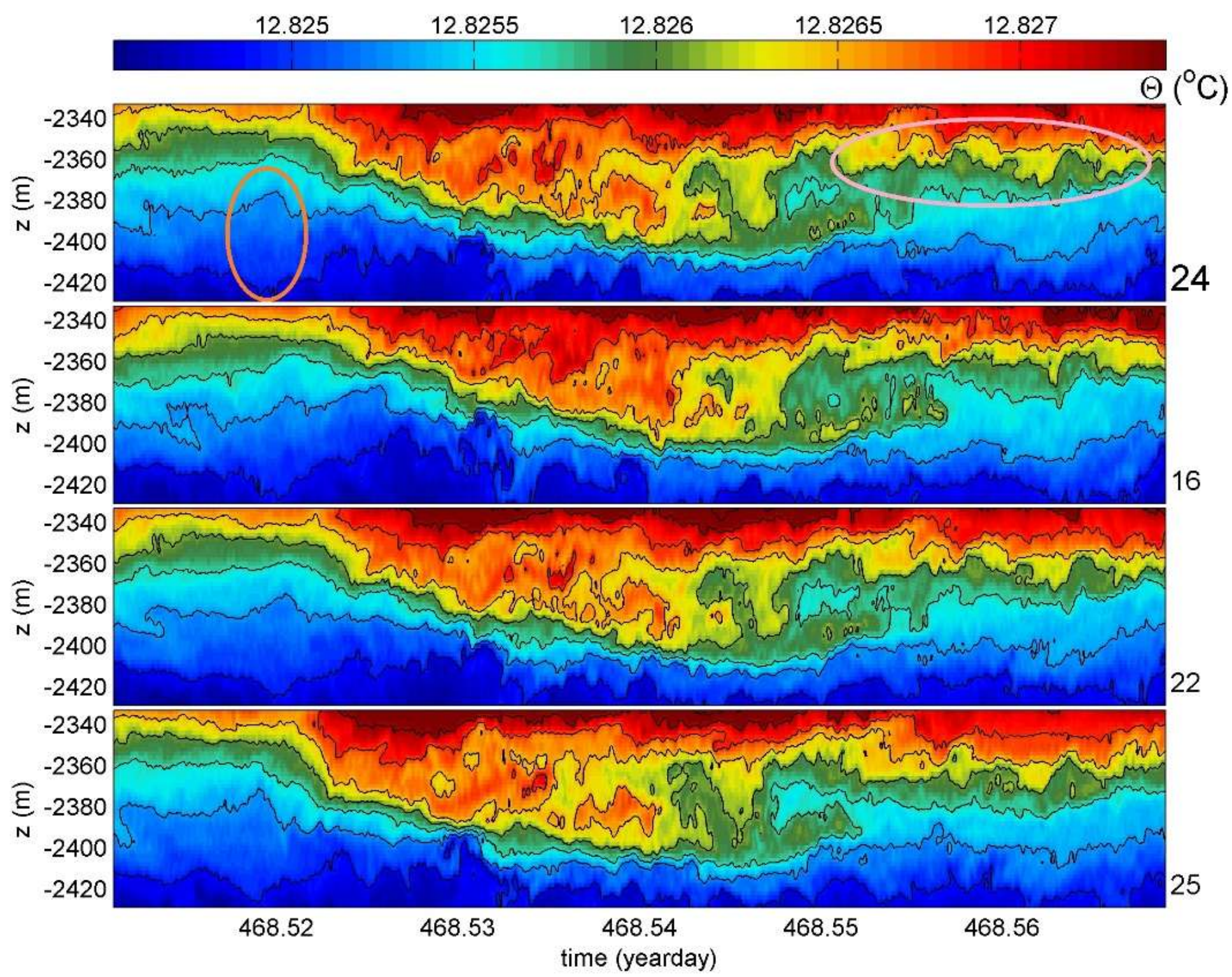

**Fig. 4.** Time-depth enlargement of one large overturn from the movie in Fig. 3e, for reference line 24 and its three neighbours. Throughout the images widening and shrinking is observed of black isotherms, which are drawn at fixed intervals of 0.0004°C. The observations evidence deep-sea turbulence generation via the process of parametric instability. Not only is the large breaker in the middle differently developing at different lines, but also the isotherm widening at all scales throughout the panel, e.g., in the ellipses (see text for explanation).

Supporting Information for

**Whipped and mixed warm clouds in the deep sea**

by Hans van Haren*, KM3NeT collaboration*

*Corresponding author: hans.van.haren@nioz.nl; km3net-pc@km3net.de

[ar]NIOZ (Royal Netherlands Institute for Sea Research), PO Box 59, Den Burg, Texel, 1790 AB, the Netherlands

**Complete author list can be found in the Appendix

**Contents of this file**

**Materials and Methods**

To study the generation and development of deep-sea turbulence, the array of sensors was distributed as a 3D-grid in nearly half a cubic hectometer and measured the temperature at 0.5 Hz using 2925 self-contained high-resolution NIOZ4 T-sensors (van Haren et al., 2021). The sensors were taped at 2-m intervals to 45 vertical lines of 125 m length. Each line was tensioned by a 1.3-kN buoy on top. Three buoys held a single-point Nortek AquaDopp current meter that measured waterflow and relative acoustic amplitude at a rate of once per 600 s. At 9.5-m horizontal intervals, the lines were attached in compacted form on small rings (Fig. 2 in van Haren et al., 2021) to crossings of a steel cable grid (Fig. S1). The grid was tensioned inside a 70 m diameter steel-tube ring, which functioned as a floating device when the mooring was towed to position under calm sea conditions. By opening valves of the steel tubes it sank to the seafloor controlled by a drag parachute. At the seafloor, the steel-tube ring functioned as a 140-kN anchor. After about 5 days, the vertical lines were chemically released from their small rings.

The large-ring mooring was deployed at the <1° flat and 2458-m deep seafloor of 42° 49.50′N, 6° 11.78′E, just 10 km south of the steep continental slope of the Northwestern Mediterranean Sea, in October 2020 (van Haren et al., 2021); data sampling starting at day 305.25 of the year. With the help from Irish Marine Institute Remotely Operated Vehicle (ROV) "Holland I", all vertical lines with sensors were successfully recovered in March 2024. Each line was cut off by the ROV.

NIOZ4 T-sensors (van Haren, 2018) are standalone self-contained instruments. The individual clocks are synchronised to a single standard clock via induction by sending a signal pulse through the steel lines every 4 hours; all T-sensors are sampled within 0.02 s. Of the large-ring mooring, one line missed synchronisation, possibly due to an electric cable failure. This line was maximum 30 s off from the others, which does not affect the studies presented here. All T-sensors stopped sampling when the file size of their memory card reached 30 MB. It implied that a maximum of 20 months of data was obtained for temperature-only sensors.

The T-sensors are calibrated in a custom-made laboratory bath that can hold 200 sensors in a titanium plate submerged in water with an anti-freeze agent (van Haren, 2018). Over a preset range, for the Mediterranean between 8°C and 18°C the bath is stabilised to within <0.0005°C at 1°C intervals, every 3-4 hours.

During post-processing several corrections are made to the data. Depending on time in the record, 50-150 (<6% of) T-sensors are not further considered due to electronics problems. The data at these sensor positions are linearly interpolated between neighbouring sensors. The T-sensors have a low instrumental noise level of nominally 0.00003°C, but under weakly stratified conditions, as found in the deep sea, some low-pass filtering has to be applied. Double elliptic, phase-preserving filters (Parks & Burrus, 1987) are used with cut-off frequencies between 700 and 3000 cpd (cycles per day), depending on the signal-noise level. Also, corrections have to be made for drift of typically 0.001°C month$^{-1}$ of the signal. A daily averaged vertical temperature profile is expected to be stable from a turbulent-overturning perspective. This is exploited to correct the instrumental drift by fitting a smooth n$^{th}$, usually 3$^{rd}$, order polynomial to the average vertical temperature profile (van Haren, 2022). In addition, the nearest of three periods of one hour duration with homogeneous temperature variations below instrumental noise level are used as reference. Under near-homogeneous conditions, vertical low-pass (noise) filtering of data was applied with cut-off at scales of 5-20 m to remove short-term bias.

These corrections allow for proper calculations of turbulence values using the method of reordering unstable overturning data to stable vertical profiles at each time step (Thorpe, 1977) under weakly stratified conditions. The method works properly for moored T-sensors under well-stratified conditions (van Haren & Gostiaux, 2012). Under unstable conditions, e.g. following geothermal heating (GH), when turbulent overturns exceed the spatial extent of the T-sensor array, manual verification is required with tuning (van Haren, 2025). In long-term automated calculations of turbulence values, periods with dominant GH are therefore fixed to a mean value established from manual investigation of ten one-day periods.

In order to use temperature as tracer in turbulence calculations using the reordering method (Thorpe, 1977) a density-temperature relationship is required. For this purpose a single shipborne Conductivity-Temperature-Depth 'CTD' profile was measured about 1 km from the mooring site during the deployment cruise.

The processed data from the mooring-array allowed for the construction of 3D movies to study small-scale internal wave and turbulence motions. For this the following steps were followed:

1) A real-time length of half a day of data was selected. The length was determined by dynamic range of temperature, adequate length for long-term drift removal, and close to the local inertial period.

2) A low-pass filter with cut-off at 700 or 3000 cycles per day was applied to all 2800 independent records.

3) Polynomial fits were applied to remove drift. The fits may be different between the lines, but the degree of the polynomial was kept fixed for all lines, each half-day period.

4) In case of near-homogeneous conditions the data were additionally low-pass filtered with cut-off equivalent of 20-m vertical scale, and 5-m scale for stratified-water conditions. The threshold between the two conditions was a daily mean 124-m vertical temperature difference of +0.0002°C.

As no bias correction is possible between the lines in the horizontal, the mean temperature values of each line are normalised to the same mean value for each half-day period. The processed data are displayed with the original distances directly downscaled, only the vertical dimension is downscaled by another factor of 2. This is uncommon for studies of the ocean of which the basins have an aspect ratio of 1:1000. Each T-sensor record is represented by a temperature-coloured filled circle at every time step.

The plots are saved into files that are loaded as frames in video-utility ‘VirtualDub’[#], compressed, and a 72-s movie is created – a factor of 600 faster than real-time.

The time lapse and movie length are adapted to the typically 0.02-0.05 m $s^{-1}$ waterflow speed, so that particles pass the entire array in a few seconds of movie time. The scale size O(100) m, resolved at 2 m vertically and 9.5 m horizontally, seems adequate for the aims of the investigation.

[#] https://www.virtualdub.org/ last accessed 10 October 2025

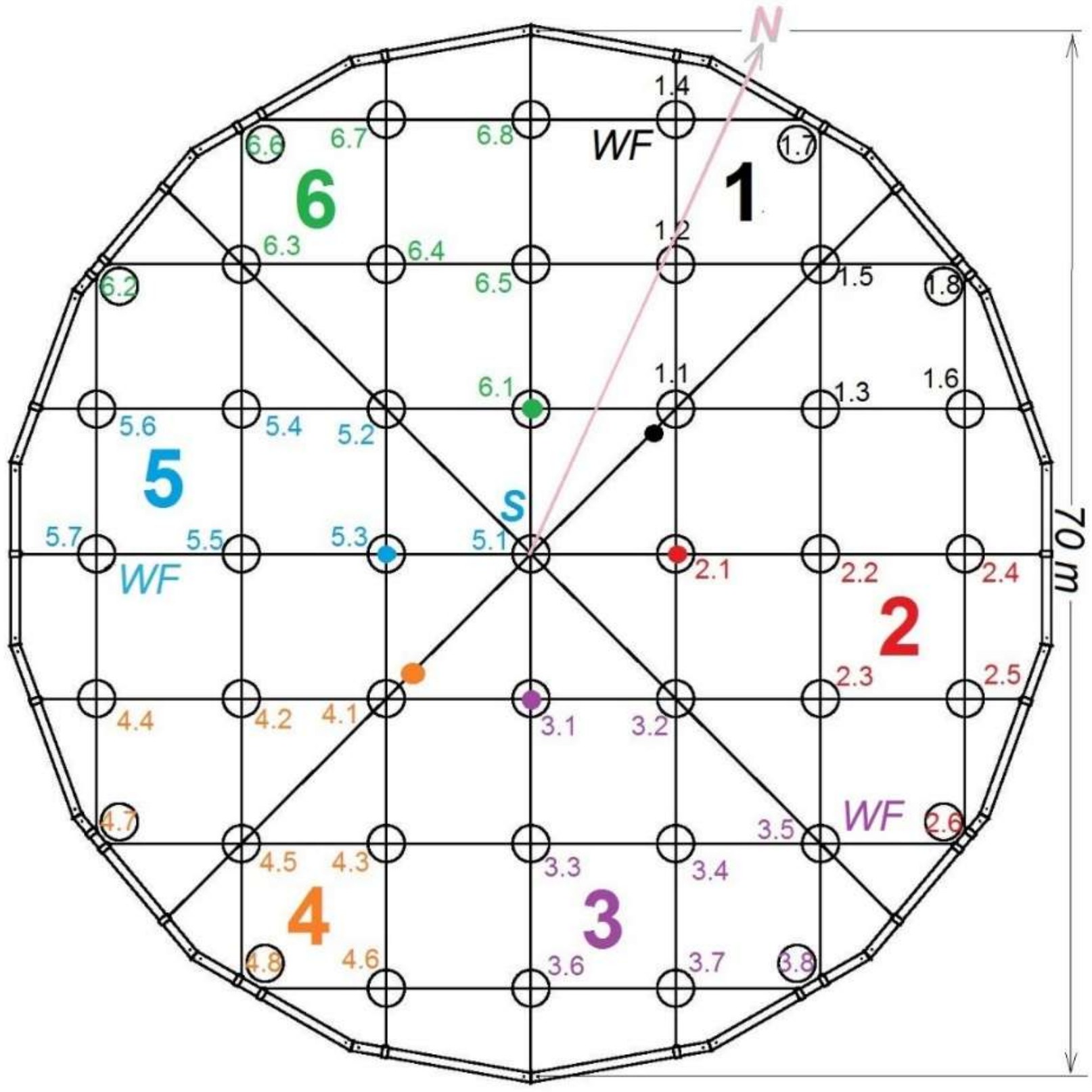

**Fig. S1.** Orientation at the seabed and layout of the large-ring mooring, with steel-cable grid and small rings holding the vertical lines at 9.5-m intervals. Lines are numbered in six synchronisation groups. Colour dots indicate group nodes and synchroniser 'S' at line 51. Here and elsewhere in the text, lines are indicate